# Two-step Dirichlet random walks

Gérard Le Caër

*Institut de Physique de Rennes, UMR UR1-CNRS 6251, Université de Rennes I, Campus de Beaulieu, Bâtiment 11A, F-35042 Rennes Cedex, France*

Tel :+33 2 23 23 53 77
Fax :+33 2 23 23 67 17

*E-mail address*:
gerard.le-caer@univ-rennes1.fr

PACS: 05.40.Fb Random walks and Lévy flights, 02.50.Cw Probability theory

Keywords:  Dirichlet random walks; random flights; Dirichlet distribution; asymmetric distributions, Pochhammer symbols, Lucas coefficients

**Abstract**

Random walks of $n$ steps taken into independent uniformly random directions in a $d$-dimensional Euclidean space $(d \geq 2)$, which are characterized by a sum of step lengths which is fixed and taken to be 1 without loss of generality, are named "Dirichlet" when this constraint is realized via a Dirichlet law of step lengths. The latter continuous multivariate distribution, which depends on $n$ positive parameters, generalizes the beta distribution $(n=2)$. It is simply obtained from $n$ independent gamma random variables with identical scale factors. Previous literature studies of these random walks dealt with symmetric Dirichlet distributions whose parameters are all equal to a value $q$ which takes half-integer or integer values. In the present work, the probability density function of the distance from the endpoint to the origin is first made explicit for a symmetric Dirichlet random walk of two steps. It is valid for any positive value of $q$ and for all $d \geq 2$. The latter pdf is used in turn to express the related density of a random walk of two steps whose step length is distributed according to an asymmetric beta distribution which depends on two parameters, namely $q$ and $q+s$ where $s$ is a positive integer.

## 1. Introduction

To model the infiltration rate of a given species into possible habitats, Pearson defined in 1905 a simple planar "random walk" (RW) which is made of a sequence of *n* steps with identical fixed lengths taken into uniformly random directions [1-2]. This idealized RW has been used recently to assess the electromagnetic compatibility of a group of *N* identical power electronic converters [3]. The cases of five to some tens of steps (*N*) were more particularly analyzed. The Pearson's RW has been applied as well to the characterization of the cosmic microwave background (CMB) [4-5]. Complex coefficients are obtained in a spherical harmonic representation of temperature maps of the CMB. Different types of random walks, associated with the spherical harmonic mode $l$, are then performed in the phase space, one being a Pearson's RW [4-5]. In addition, the analysis of the temperature and polarization of the CMB led Reimberg and Abramo [6-7] to define random flights made of two successive stages in spaces with different dimensions $d_1, d_2 \left(2 \leq d_1 \leq d_2\right)$, both with deterministic step lengths. Such flights emerge from the treatment of Boltzmann equations which codify the interplay between collisional physics and free propagation. The coefficients of a multipole decomposition of the temperature and polarization of the CMB are determined from these equations. For each stage, the space dimension is determined by the order of the multipole which dominates it [7].

Variations on the theme of Pearson's random walk involve space dimensions higher than two, changes of step length distributions, deviations of step orientations from a uniform repartition and the introduction of correlations between steps [6-34]. A frequent change consists in allowing step lengths to vary according to some continuous probability law. Such modifications find applications in diverse fields such as physics, biology, ecology ([4-7, 9–20] and references therein). A few examples, by no means exhaustive, are given hereafter. Random walks with exponentially distributed step lengths were studied in 2D by Stadje [13] as a possible description of the motion of microorganisms on planar surfaces and in 3D by Vignolles et al. [14] to model the chemical vapor infiltration method used to prepare ceramic-matrix composites. To study the relation between the Boltzmann equation and the underlying stochastic processes, Zoia et al. [15] investigated exponential flights in $\mathbb{R}^d$. The probability density of finding a particle at position *r* at the *n*-th collision was determined for an infinite medium. The cases were $d = 1, 2$ find applications respectively in the field of electron transport in nanowires or carbon nanotubes and in the study of the dynamics of chemical and biological species on surfaces. Exponential walks are quite naturally extended to random walks with gamma distributed step lengths (appendix A).

An important class of walks is that of Lévy flights whose step lengths have heavy-tailed probability distributions. Many organisms are believed to perform Lévy flights in their search for resources [16-20]. The Levy-flight foraging hypothesis is the subject of much debate but there is presently a growing consensus that many organisms diffuse anomalously. Alternative models to Lévy flights or the emergence of Levy patterns from composite models, are regularly put forward and discussed controversially (see for instance [18]). We notice that a shifted gamma distribution, with a small shape factor ~0.3, was used for some time to account for flight durations of sea birds [17].

The applications of the previous random walks imply neither that the walk space is a physical space with a dimension of at most three nor that the step length distributions are limited to a few standard mathematical forms.

The present work focuses on random walks in a $d$-dimensional Euclidean space $(d \geq 2)$ whose step lengths have a Dirichlet distribution. The Dirichlet distribution is applied for instance to model fragmentation or compositional data [35]. Further, gamma and Dirichlet distributions are strongly connected as are the associated random walks (appendix A).

*1.1 Dirichlet random walks* of $n$ steps

Random walks of $n$ steps in $\mathbb{R}^d$ $(d \geq 2)$ taken into independent uniformly random directions, with unequal step lengths $l_k$ $(k=1,..,n)$ which obey the additional constraint of a constant total sum $\sum_{k=1}^{n} l_k = S = cst$, were investigated recently [21-29]. The constant $S$ is taken hereafter as equal to 1 without loss of generality (eq. 5, section 2). The problem of the step length distribution of such walks is thus directly related to the broken stick problem, i.e. the problem of the random splitting of a unit interval. The Dirichlet distribution has been more particularly considered as an appropriate step length distribution [22-29]. Following Letac and Piccioni [29], the associated random walks are named hereunder "Dirichlet random walks" and referred to as $W(d, n, \boldsymbol{q}_{(n)})$. The Dirichlet distribution of the random vector $\boldsymbol{L}_{(n)} = (L_1, L_2, ..., L_n)$ $\left( L_n = 1 - \sum_{k=1}^{n-1} L_k \right)$, denoted here as $D(\boldsymbol{q}_{(n)})$, with parameters $\boldsymbol{q}_{(n)} = (q_1, ..., q_n)$ has a multivariate probability density function (pdf) given by ([36], p. 18):

$$\begin{cases} f(l_1,..,l_m) = K_n \prod_{i=1}^{n} l_i^{q_i-1} \\ l_n = 1 - \sum_{i=1}^{m} l_i, \quad l_i > 0, \ i = 1,..,n \end{cases} \tag{1}$$

where $m = n-1$, $K_n = \Gamma(n\bar{q}) \Big/ \left( \prod_{i=1}^{n} \Gamma(q_i) \right)$, $n\bar{q} = \sum_{i=1}^{n} q_i$. For $n = 2$, the Dirichlet distribution reduces to a beta distribution denoted hereafter as $Be(q_1, q_2)$ (eq. 2). For convenience, these two-step beta random walks will still be named "Dirichlet". In the literature, special attention has been paid to "symmetric" Dirichlet random walks, $W(d,n,q)$, i.e. to walks whose step lengths are distributed according to a symmetric Dirichlet distribution for which $\boldsymbol{q}_{(n)} = (q,q,...,q)$. As the notation $W(d,n,q)$ is self-explanatory, the word "symmetric" is omitted to designate them. There is a close connection between a symmetric Dirichlet random walk $W(d,n,q)$ and a gamma random walk $G(d,n,q)$ whose step lengths have identical and independent gamma distributions $\gamma(q,1)$ (appendix A). The pdf of the endpoint position of $G(d,n,q)$ is obtained from a single integral from the pdf of the endpoint position of $W(d,n,q)$ (eq. A.8, [25-26]).

One of the initial motivations for studying constrained exponential random walks, $W(d,n,1)$, was to answer the question as to whether it is possible to find triplets $(d,n,1)$ for which the endpoints are uniformly distributed on the unit ball of $\mathbb{R}^d$ [21]. The latter quest was extended to walks $W(d,n,q)$ [25] and more generally to hyperuniform random walks where the word "hyperuniform" used in [29] is preferred here to the term "hyperspherical uniform" used in [25-26]. A $n$-step random walk in $\mathbb{R}^d$ is said to be hyperuniform of type $k > d$ if the distribution of the endpoint of the walk in $\mathbb{R}^d$ is identical with the distribution of the projection in the walk space of a point uniformly distributed on the surface of the unit hypersphere of $\mathbb{R}^k$ [25-26,29]. The pdf of the position $\boldsymbol{R}$ of the endpoint of a Dirichlet random walk and that of the distance $R = \|\boldsymbol{R}\|$ from the endpoint to the starting point will be denoted respectively as $p_{\boldsymbol{q}_{(n)}}^{(d)}(\boldsymbol{r})$ and $P_{\boldsymbol{q}_{(n)}}^{(d)}(r)$. These two pdf's, which are simply related (eq. 4), can be considered interchangeably. For hyperuniform random walks, $p_{\mathbf{q}_{(n)}}^{(d)}(r)$ is $\propto \left(1-r^2\right)^{(k-d-2)/2}$. A uniform random walk on the unit ball of $\mathbb{R}^d$ defined above is thus hyperuniform of type $k = d+2$.

The pdf's $p_{\mathbf{q}_{(n)}}^{(d)}(r)$ were derived for any space dimension $d \geq d_0$ and any $n \geq 2$ for the three following families of walks $W(d,n,q)$: $q = d/2 - 1$ with $d_0 = 3$, $q = d-1$ and $q = d$, both with $d_0 = 2$ [25-26]. The walks with $q = d/2 - 1$ and $q = d - 1$ were shown to be hyperuniform of type $k = n(d-2)+2$ and $k = n(d-1)+1$ respectively [25]. In addition, the pdf's $p_{\mathbf{q}_{(n)}}^{(d)}(r)$ were obtained for particular values of two of the three parameters of the triplet $(d,n,q)$ such as $(d,2,1)$ [22] and $(6,n,1)$ [24] or for particular values of these three parameters. Altogether, $q$ takes either half-integer or integer values in all cases where $p_{\mathbf{q}_{(n)}}^{(d)}(r)$ has been made explicit.

## 1.2 The focus on two-step Dirichlet random walks

It is much simpler to study two-step Dirichlet random walks $W(d,2,q)$ than the general Dirichlet random walks mentioned above so that the question of their relevance might even arise. However, their unequalled advantage lies in the existence of an explicit expression of the pdf of the endpoint distance, $P_{\mathbf{q}_{(2)}}^{(d)}(r) \left( \equiv P_{q,q}^{(d)}(r) \right)$, which is valid for any value of $q > 0$ and for any $d \geq 2$ such that $q(d) > 0$. To the best of our knowledge, the latter expression has not yet been given explicitly in the literature.

In the present work, we obtain first the density $P_{q,q}^{(d)}(r)$ (section 3). Second, the relevance and the usefulness of the permutation invariant distribution associated with an asymmetric step length distribution are discussed (section 4). The former distribution is used among others to establish that the two following families of Dirichlet random walks $W(d,n,\mathbf{q}_{(n)})$, where $\mathbf{q}_{(n)}$ is respectively $(q,q,...,q)$ and $(q+1,q,...,q)$, are indistinguishable for any $q(d) > 0$ (section 4). Last, the results of the two previous steps are applied to the derivation of the pdf $P_{q+s,q}^{(d)}(r)$ of two-step random walks $W(d,2,\boldsymbol{q}_{(2)}) \equiv W(d,2,(q+s,q))$ which depend on two different Dirichlet parameters, $q>0$ and $q+s$ where $s$ is a positive integer (section 5). A second formal representation of $P_{q+s,q}^{(d)}(r)$ is derived. Both representations are shown to be equivalent by a method of moments.

## 2. Notations

As usual upper-case letters will be used to denote random variables (r.v.) and lower-case letters for the values they take. The mean of a function $f(X)$ of a continuous random variable $X$, with a pdf $p_X(x)$ whose support is $D$, will be denoted hereafter as $\langle f(X) \rangle = \int_D f(x) p_X(x) dx$.

The Pochhammer symbol $(a)_k$, its duplication formula [37], the beta function $B(\alpha,\beta)$, the beta distribution, $L \sim Be(q_1,q_2)$, whose pdf is $f_L(l)$ and its moments $\langle L^k \rangle$, will be used repeatedly throughout the text:

$$\begin{cases} (a)_k = \Gamma(a+k)/\Gamma(a) = a(a+1)..(a+k-1),\ (a)_0 = 1 \\ (a+b)_{2k} = 4^k \left(\dfrac{a+b}{2}\right)_k \left(\dfrac{a+b+1}{2}\right)_k \\ B(\alpha,\beta) = \Gamma(\alpha)\Gamma(\beta)/\Gamma(\alpha+\beta) = \int_0^1 l^{\alpha-1}(1-l)^{\beta-1}\,dl \\ L \sim Be(q_1,q_2) \Rightarrow f_L(l) = \dfrac{l^{q_1-1}(1-l)^{q_2-1}}{B(q_1,q_2)} \quad l \in [0,1] \\ \langle L^k \rangle = B(q_1+k,q_2)/B(q_1,q_2) = (q_1)_k/(q_1+q_2)_k \end{cases} \quad (2)$$

where $\Gamma(x)$ is the classical gamma function, $\alpha,\beta>0$, $q_1,q_2>0$ and $k$ is taken here to be a positive integer.

The position of the endpoint of a Dirichlet random walk $W(d,n,\boldsymbol{q}_{(n)})$ is $(n \geq 2)$:

$$\boldsymbol{R} = \sum_{i=1}^{n} L_i \boldsymbol{U}_i^{(d)} \qquad (3)$$

where the $\boldsymbol{U}_i^{(d)}$ are $n$ independent unit vectors uniformly distributed over the surface of the hypersphere in $\mathbb{R}^d$. The $L_i$ $(i=1,..,n)$ follow a Dirichlet law $D(\boldsymbol{q}_{(n)})$. For the spherically symmetric walks described by eq. 3, the pdf of the endpoint position $\boldsymbol{R}$ and the pdf of the endpoint distance $R = \|\boldsymbol{R}\|$ of a $W(d,n,\boldsymbol{q}_{(n)})$ walk depend only on $r = \|\boldsymbol{r}\| > 0$. Both are related through:

$$P_{\boldsymbol{q}_{(n)}}^{(d)}(r) = \frac{2\pi^{d/2}}{\Gamma(d/2)} r^{d-1} p_{\boldsymbol{q}_{(n)}}^{(d)}(r) \qquad (4)$$

In the following, we will most often consider the pdf $P_{\boldsymbol{q}_{(n)}}^{(d)}(r)$. As we will deal essentially with two-step walks, the components $(q_1, q_2)$ of $\boldsymbol{q}_{(2)}$ will be given explicitly in the notations. The position and distance pdf's associated with any walk $W(d,2,(q_1,q_2))$ will respectively be denoted as $p_{q_1,q_2}^{(d)}(\boldsymbol{r})$ and $P_{q_1,q_2}^{(d)}(r)$. Finally, once pdf's are obtained for a walk $W(d,n,\boldsymbol{q}_{(n)})$ of a total length of 1, then densities $p_{\boldsymbol{q}_{(n)}}^{(d,S)}(\vec{r})$ and $P_{\boldsymbol{q}_{(n)}}^{(d,S)}(r)$ are immediately calculated for an arbitrary total walk length $S$ from:

$$\begin{cases} p_{\boldsymbol{q}_{(n)}}^{(d,S)}(\boldsymbol{r}) = \dfrac{1}{S^d} p_{\boldsymbol{q}_{(n)}}^{(d)}\left(\dfrac{\boldsymbol{r}}{S}\right) \\ P_{\boldsymbol{q}_{(n)}}^{(d,S)}(r) = \dfrac{1}{S} P_{\boldsymbol{q}_{(n)}}^{(d)}\left(\dfrac{r}{S}\right) \end{cases} \qquad (r \in [0,S]) \qquad (5)$$

Similar relations were given previously as eq. 16 of [25] and as eq. 2 of [26] where $S$ is replaced by $l$. In the second relation of each equation, $l^d$ must be replaced by $l$. The results discussed in [25-26] are not affected by this error as they were obtained from the first part of eq. 5 followed by the application of eq. 4.

## 3. Two-step Dirichlet random walks $W(d,2,q)$

The final position of a walk $W(d,2,q)$ is more simply written as (eq. 3):

$$\mathbf{R} = L\mathbf{U}_1^{(d)} + (1-L)\mathbf{U}_2^{(d)} \qquad (6)$$

From eq. 6, we obtain the square of the final distance $R^2 = \mathbf{R}.\mathbf{R}$:

$$R^2 = 1 - 2L(1-L)\left(1 - \mathbf{U}_1^{(d)}.\mathbf{U}_2^{(d)}\right) \qquad (7).$$

Defining:

$$\begin{cases} Y = 4L(1-L) \\ X = 1-R^2 \\ \cos\Theta = \mathbf{U}_1^{(d)}.\mathbf{U}_2^{(d)} \\ Z = (1-\cos\Theta)/2 = \sin^2(\Theta/2) \end{cases} \qquad (8)$$

eq. 7 becomes:

$$X = YZ \qquad (9)$$

In eq. 9, the r.v.'s $Y$ and $Z$ are independent and have beta distributions, respectively $Y \sim Be\left(q, \frac{1}{2}\right)$ when $L \sim Be(q,q)$ and $Z \sim Be\left(\frac{d-1}{2}, \frac{d-1}{2}\right)$ [29]. Appendix B gives a derivation of the distribution of $Y$ for $L \sim Be(q_1,q_2)$. The latter distribution is relevant for the discussion presented in section 4. The distribution of $Z$ is simply found to be $Z \sim Be\left(\frac{d-1}{2}, \frac{d-1}{2}\right)$ from the pdf of the polar angle $\Theta$ (see for instance [36] p. 104), $p_\Theta(\theta) = \sin^{d-2}\theta / B(1/2, (d-1)/2)$. The moments of $Z$ are thus $\langle Z^k \rangle_d = ((d-1)/2)_k / (d-1)_k$ (eq. 2). The moments of $Y$, $\langle Y^k \rangle_{q_1,q_2} = \left\langle [4L(1-L)]^k \right\rangle_{q_1,q_2}$, are in turn readily obtained from eq. 2 for a walk $W(d,2,(q_1,q_2))$ whose length distribution is then $L \sim Be(q_1,q_2)$. Eq. 9 yields finally:

$$\langle X^k \rangle_{d,q_1,q_2} = \left\langle (1-R^2)^k \right\rangle_{d,q_1,q_2} = \langle Y^k \rangle_{q_1,q_2} \langle Z^k \rangle_d = 4^k \times \frac{(q_1)_k (q_2)_k}{(q_1+q_2)_{2k}} \times \frac{((d-1)/2)_k}{(d-1)_k} \qquad (10)$$

The pdf of a r.v., which is a product of two independent beta r.v.'s, is explicitly known as a function of the parameters of the two beta laws ([38-41] and appendix C). With $Y \sim Be\left(q, \frac{1}{2}\right)$ and $Z \sim Be\left(\frac{d-1}{2}, \frac{d-1}{2}\right)$, the pdf of $X$ writes (eq. C.1):

$$p_X(x) = \frac{2^{d-2}}{B(q,1/2)} \times x^{q-1}(1-x)^{(d-2)/2} \,_2F_1\left(q+1-\frac{d}{2}, \frac{d-1}{2}; \frac{d}{2}; 1-x\right) \quad (x \in [0,1]) \tag{11}$$

The common support of the distributions of $X$ and of $R$ is [0,1]. As $r$ is a monotonically decreasing function of $x$, $r = \sqrt{1-x}$, the pdf of $R$ is obtained from the pdf $p_X(x)$ to be $P_{q,q}^{(d)}(r) = 2r p_X(1-r^2)$. From eq. 11, we get then $(d \geq 2, q > 0, r \in [0,1])$:

$$P_{q,q}^{(d)}(r) = \frac{2^{d-1}}{B(q,1/2)} \times r^{d-1}(1-r^2)^{q-1} \,_2F_1\left(q+1-\frac{d}{2}, \frac{d-1}{2}; \frac{d}{2}; r^2\right) \tag{12}$$

Transforming the Gauss hypergeometric function [42], eq.12 can be written equally as:

$$P_{q,q}^{(d)}(r) = \frac{2^{d-1}}{B(q,1/2)} \times r^{d-1}(1-r^2)^{(d-3)/2} \,_2F_1\left(d-1-q, \frac{1}{2}; \frac{d}{2}; r^2\right) \tag{13}$$

The associated pdf's of the endpoint position, $p_{q,q}^{(d)}(r)$, are readily obtained from eq. 4.

Previously known pdf's are retrieved from eq. 12 or from eq. 13, adding eq. 4 if necessary. For instance, the pdf $p_{1,1}^{(d)}(r)$ was given in [22]. The specific parameters $q = d/2 - 1$ and $q = d - 1$ of the two families of hyperuniform Dirichlet random walks mentioned in section 1.1 are seen to appear quite naturally when the first arguments of the hypergeometric functions of eq. 12 and of eq. 13 are made equal to zero. We obtain:

$$P_{q,q}^{(d)}(r) = \begin{cases} \frac{2^{d-1}}{B(d/2-1,1/2)} \times r^{d-1}(1-r^2)^{(d-4)/2} & (q = d/2-1,\ d \geq 3) \\ \frac{2^{d-1}}{B(d-1,1/2)} \times r^{d-1}(1-r^2)^{(d-3)/2} & (q = d-1,\ d \geq 2) \\ \frac{2^{-d-1}}{B(d,d+1)} \times \frac{r^{d-1}}{d} \times (d-r^2)(1-r^2)^{(d-3)/2} & (q = d,\ d \geq 2) \end{cases} \tag{14}$$

These pdf's agree with those of table 1 of [25] for $n=2$, $q=d/2-1, d-1$ and with eq. 44 of [26] for $n=2$, $q=d$. Finally, eqs 4 and 12 give $p_{1,1}^{(6)}(r) = (8/3\pi^3)(6-5r^2)$ in agreement with eq. 13 of [24] for $S = ct = 1$.

In addition, we performed Monte-Carlo simulations of $W(d,2,q)$ random walks for arbitrary values of $q$. Figure 1 $(s=0)$ gives an example, among many others, of a comparison between simulated and calculated results for $q = 15/4$.

In the following, we will use the pdf $P_{q,q}^{(d)}(r)$ determined above, to express the corresponding pdf's for the random walks, $W(d,2,(q+s,q))$, where $s$ is a positive integer. The independent step length of a walk $W(d,2,(q+s,q))$ is distributed according to asymmetric beta distributions, $L \sim Be(q+s,q)$ (eq. 2, fig. 1). Before calculating the abovementioned pdf's, we first discuss the consequences of the use of such an asymmetric distribution.

## 4. Asymmetric distributions of step lengths

*4.1 Simple considerations and their consequences*

We consider a $n$-step random walk in $\mathbb{R}^d$ whose final position (eq. 3) is rewritten again as:

$$\boldsymbol{R} = \sum_{i=1}^{n} L_i \boldsymbol{U}_i^{(d)} \equiv \sum_{i=1}^{n} L_{\sigma(i)} \boldsymbol{U}_{\sigma(i)}^{(d)} \quad (15)$$

where the sole constraint is still, $\sum_{i=1}^{n} L_i = 1$. In eq. 15, $\sigma = (\sigma(1), \sigma(2), ..., \sigma(n))$ denotes any of the $n!$ permutations of $(1, 2, ..., n)$. An asymmetric distribution of the $L_i$'s means here that the multivariate distribution of lengths $f_n(l_1, l_2, ..., l_{n-1}) \left(l_n = 1 - \sum_{i=1}^{n-1} l_i\right)$ is not invariant under permutations of the $l_i$'s. Thus, the univariate marginal distributions are not all identical. When convenient, the multivariate distribution, which is not necessarily a Dirichlet distribution, will equally be written hereafter as $f(l_1, l_2, ..., l_n)$. The sum which gives $\boldsymbol{R}$ is commutative (eq. 15). The $n!$ possible attributions of a set of lengths $(l_k, k=1,..,n)$ to the steps numbered $1,2,..,n$ result then in undistinguishable walks. Then, the principle of indifference states that each permutation should be given a probability of $1/n!$ with the consequence that all steps end up with identical length distributions independently of their (arbitrary) order. It suffices then to symmetrize the initial distribution of $L$ to make it invariant under permutations. We conclude that it is this permutation invariant distribution which is the sole meaningful step length distribution. An example is discussed in appendix B for $n=2$. The permutation invariant distribution associated with $f(l_1, l_2,..., l_n)$ $(\equiv f_n(l_1, l_2,..., l_{n-1}))$ is:

$$f_n^{(\Sigma)}(l_1, l_2,..., l_{n-1}) = \frac{1}{n!} \sum_\sigma f\left(l_{\sigma(1)}, l_{\sigma(2)},..., l_{\sigma(n)}\right) \tag{16}$$

where the sum runs over the $n!$ permutations $\sigma$ of $(1,2,..,n)$. Constrained random walks with different initial length distributions are thus undistinguishable if and only if the distributions $f_n^{(\Sigma)}(l_1, l_2,..., l_{n-1})$ associated with them are identical.

We apply now the previous discussion to random walks whose step length distributions are asymmetric Dirichlet distributions in the above sense. Dirichlet random walks with such step length distributions are not really "Dirichlet" as the relevant distribution (eq. 16) is no more a symmetric Dirichlet distribution, except in some particular cases (eq. 18 below with $q_1 = q+1$), but a mixture of asymmetric Dirichlet distributions. To distinguish between these two types of walks, the latter will be designated in abridged form as "asymmetric Dirichlet random walks".

*4.2 Application to some asymmetric Dirichlet random walks*

We consider more particularly Dirichlet distributions of step lengths whose parameters are all equal except one taken to be the first, $\boldsymbol{q}_{(n)} = (q_1, q,..,q)$. The associated pdf (eq. 1) is explicitly:

$$\begin{cases} f(l_1, l_2, ..., l_m) = \dfrac{\Gamma((n-1)q + q_1)}{\Gamma(q_1)[\Gamma(q)]^{n-1}} l_1^{q_1-1} \prod_{i=2}^{n} l_i^{q-1} \\ l_n = 1 - \sum_{i=1}^{n} l_i \qquad l_i > 0 \ \forall i \end{cases} \qquad (17)$$

The marginal distributions of all $L_k$ $(k>1)$ are identical beta distributions $L_k \sim Be(q, mq+1)$ while that of $L_1$ is $Be(q+1, mq)$ ([36] and eq. A.3). The permutation invariant distribution associated with eq. 17 is then:

$$f_n^{(\Sigma)}(l_1, l_2, ..., l_{n-1}) = \dfrac{1}{n} \times \dfrac{\Gamma((n-1)q + q_1)}{\Gamma(q_1)[\Gamma(q)]^{n-1}} \times \left[\prod_{i=1}^{n} l_i^{q-1}\right] \times \left[\sum_{i=1}^{n} l_i^{q_1-q}\right] \qquad (18)$$

as all $l_i$, except one, play the same role. Thus, the permutation invariant density (eq. 18) reduces to the Dirichlet law whose parameters are all equal to $q$ when $q_1 = q+1$ because $\sum_{i=1}^{n} l_i = 1$. In other words, the asymmetric Dirichlet random walk whose parameters are $(q+1, q, ..., q)$ is identical with the symmetric Dirichlet random walk whose parameters are $(q, q, ..., q)$. The latter conclusion is valid for any $q > 0$. This results in the identity of the pdf's, $p_{q+1,q,..q}^{(d)}(r) \equiv p_{q,q,..q}^{(d)}(r)$ and $P_{q+1,q,..q}^{(d)}(r) \equiv P_{q,q,..q}^{(d)}(r)$ for any value of $n \geq 2$. In section 5 of [25], it was shown that the $n$-steps Dirichlet random walks in $\mathbb{R}^d$, whose parameters are $(q, q, ..., q)$ and $(q+1, q, ..., q)$ $(q = d/2 - 1, d-1)$, yield both hyperuniform random walks of type $k$ where $k$ is $n(d-2)+2$ for $q = d/2-1$ and $n(d-1)+1$ for $q = d-1$. The families $(q, q, ..., q)$ and $(q+1, q, ..., q)$ were incorrectly considered to be different [25].

## 5. Asymmetric Dirichlet random walks of two steps $W(d, 2, (q+s, q))$

In the case of two steps, now with $q_1 = q + s$ where $q$ takes any positive value and $s$ is a positive integer, the permutation invariant distribution (eq. 18) writes:

$$f_2^{(\Sigma)}(l) = \frac{[l(1-l)]^{q-1}}{2B(q+s,q)} \times [l^s + (1-l)^s] \qquad (19)$$

Two formal representations of the pdf $P_{q+s,q}^{(d)}(r)$ are obtained from this permutation invariant pdf by the simple methods described below. The first representation is derived from eq. 9, $X = YZ$ (section 5.1) while the second (section 5.2) is based on an expansion of $l^s + (1-l)^s$ in terms of powers of $l(1-l)$.

*5.1 First representation of the pdf $P_{q+s,q}^{(d)}(r)$*

In appendix B, the distribution of $Y = 4L(1-L)$ (eq. 8) is shown to be a weighted sum of $1 + \lfloor \frac{s}{2} \rfloor$ beta distributions with weights $c_k = \binom{s}{2k} \times \frac{B(q, k+1/2)}{2^{2q+s-1} B(q+s, q)}$ $\left(k = 0, ..., \lfloor \frac{s}{2} \rfloor\right)$ (eq. B.6). The distribution of $X = YZ = 1 - R^2$ (eqs 8 and 9) is consequently a mixture of $1 + \lfloor \frac{s}{2} \rfloor$ distributions of r.v.'s $X_k$, $p_X(x) = \sum_{k=0}^{\lfloor s/2 \rfloor} c_k p_{X_k}(x)$. The distribution of each $X_k$ is that of a product of independent beta r.v.'s, $X_k = Y_k Z$, with $Y_k \sim Be(q, k+1/2)$ (eq. B.6) and $Z \sim Be\left(\frac{d-1}{2}, \frac{d-1}{2}\right)$ (section 3). It is given by the following relation (eq. C.1):

$$\begin{cases} p_{X_k}(x) = \alpha_k x^{q-1} (1-x)^{k+d/2-1} \times {}_2F_1\left(q+k+1-\frac{d}{2}, \frac{d-1}{2}; \frac{d}{2}+k; 1-x\right) \\ \alpha_k = \dfrac{B((d-1)/2, k+1/2)}{B(q, k+1/2) B((d-1)/2, (d-1)/2)} \end{cases} \qquad (20)$$

The sought-after pdf is finally obtained from the relation, $P_{q+s,q}^{(d)}(r) = 2r p_X(1-r^2)$ (section 3):

$$\begin{cases} P_{q+s,q}^{(d)}(r) = \alpha_{q+s,q}^{(d)} r^{d-1} (1-r^2)^{q-1} \times \left\{ \sum_{k=0}^{\lfloor s/2 \rfloor} \omega_k r^{2k} \times {}_2F_1\left(q+k+1-\frac{d}{2}, \frac{d-1}{2}; \frac{d}{2}+k; r^2\right) \right\} \\ \alpha_{q+s,q}^{(d)} = \dfrac{1}{2^{2q+s-2} B(q+s,q) B((d-1)/2, (d-1)/2)} \quad , \quad \omega_k = \binom{s}{2k} \times B((d-1)/2, k+1/2) \end{cases} \quad (21)$$

This first representation is a sum of $1 + \left\lfloor \dfrac{s}{2} \right\rfloor$ non-negative contributions. This is not the case for the second one which is now derived.

*5.2 Second representation of the pdf $P_{q+s,q}^{(d)}(r)$*

To express explicitly $f_2^{(\Sigma)}(l)$ (eq. 19) as a linear combination of symmetric pdf's, we need first to expand $l^s + (1-l)^s$ in terms of powers of $l(1-l)$. For this end, we use a classical expansion [43-44]:

$$x^s + y^s = \sum_{j=0}^{\lfloor s/2 \rfloor} (-1)^j C(s,j) (xy)^j (x+y)^{s-2j} \qquad (22)$$

For reasons made clear later (eq. 32, appendix D), we prefer to write it as:

$$\frac{x^s + y^s}{(x+y)^s} = \sum_{j=0}^{\lfloor s/2 \rfloor} (-1)^j C(s,j) \frac{(xy)^j}{(x+y)^{2j}} \qquad (23)$$

$(x+y \neq 0)$ and $s \geq 2$. The case where $s=1$, which reduces to $1=1$, is solved in section 4.2 with the result that $P_{q+1,q}^{(d)}(r) \equiv P_{q,q}^{(d)}(r)$. This case is no longer considered even if eqs 26 and 28 below hold for $s=1$. The numbers $C(s,j)$, known as coefficients of Lucas (or Cardan) polynomials [45], are given by:

$$C(s,j) = \frac{s}{(s-j)}\binom{s-j}{j} \left(0 \leq j \leq \left\lfloor \frac{s}{2} \right\rfloor\right) = \begin{cases} 1 & \text{if } j=0 \\ \dfrac{s}{j}\binom{s-j-1}{j-1} & \left(1 \leq j \leq \left\lfloor \frac{s}{2} \right\rfloor\right) \end{cases} \qquad (24)$$

From eq. 22, we write:

$$l^s + (1-l)^s = \sum_{j=0}^{\lfloor s/2 \rfloor} (-1)^j C(s,j) \left[l(1-l)\right]^j \quad (s \geq 2) \quad (25)$$

We are now ready to express explicitly $f_2^{(\Sigma)}(l)$ as a linear combination of $1+\left\lfloor \dfrac{s}{2} \right\rfloor$ beta distributions from which the pdf $P_{q+s,q}^{(d)}(r)$ can be obtained immediately from the known pdf $P_{q',q'}^{(d)}(r)$ given in section 3 (eqs 12 and 13). First, eq. 19 writes:

$$f_2^{(\Sigma)}(l) = \sum_{j=0}^{\lfloor s/2 \rfloor} (-1)^j w_j \left\{ \frac{[l(1-l)]^{q+j-1}}{B(q+j, q+j)} \right\} \quad (s \geq 2) \quad (26)$$

With:

$$w_j = \frac{C(s,j)}{2} \times \frac{B(q+j, q+j)}{B(q+s, q)} = \frac{C(s,j)}{2} \times \frac{(2q)_s (q)_j (q)_j}{(q)_s (2q)_{2j}} \quad (27)$$

We deduce then the sought-after pdf of $R$:

$$P_{q+s,q}^{(d)}(r) = \sum_{j=0}^{\lfloor s/2 \rfloor} (-1)^j w_j P_{q+j,q+j}^{(d)}(r) \quad (q>0, s \geq 2) \quad (28)$$

The coefficients $w_j$ of this expansion, which depend only on the distribution of $L$, remain the same for any space dimension $d$. From eq. 13 (noticing that $\dfrac{2^{d-2q'}}{B(q',q')} = \dfrac{2^{d-1}}{B(q',1/2)}$), eq. 28 can be written explicitly as:

$$P_{q+s,q}^{(d)}(r) = \frac{2^{d-2q-1} r^{d-1} (1-r^2)^{(d-3)/2}}{B(q+s,q)} \times \left\{ \sum_{j=0}^{\lfloor s/2 \rfloor} \frac{(-1)^j C(s,j)}{4^j} \times {}_2F_1\left(d-1-q-j, \frac{1}{2}; \frac{d}{2}; r^2\right) \right\} \quad (29)$$

Equivalently, from eq. 12:

$$P_{q+s,q}^{(d)}(r) = \frac{2^{d-2q-1}}{B(q+s,q)} \times r^{d-1}(1-r^2)^{q-1} \times$$

$$\times \left\{ \sum_{j=0}^{\lfloor s/2 \rfloor} \frac{(-1)^j C(s,j)}{4^j} \times (1-r^2)^j \times {}_2F_1\left(q+j+1-\frac{d}{2}, \frac{d-1}{2}; \frac{d}{2}; r^2\right) \right\} \quad (30)$$

A direct proof of the identity of the two representations of the pdf $P_{q+s,q}^{(d)}(r)$ might be obtained from a cascade of Gauss' transformations between contiguous hypergeometric functions applied to eq. 21 as done in appendix E for $s = 2,3$. Indeed, the third argument $\frac{d}{2}+k'$ of the hypergeometric functions ${}_2F_1\left(q+k'+1-\frac{d}{2}, \frac{d-1}{2}; \frac{d}{2}+k'; r^2\right)$ of eq. 21 must be transformed into $\frac{d}{2}$ for any $k'$ $\left(k'=1,...,\left\lfloor \frac{s}{2} \right\rfloor\right)$. However, this calculation appears to be complicated. As distributions with bounded supports are uniquely determined by their moments of positive integer order $k$, it suffices to prove that both representations yield identical moments $\langle X^k \rangle_{d,q+s,q} = \langle (1-R^2)^k \rangle_{d,q+s,q}$ (eq. 8) for the walks $W(d,2,(q+s,q))$.

Indeed, the pdf of $R$ is obtained from the pdf $p_X(x)$ as $P_{q+s,q}^{(d)}(r) = 2rp_X(1-r^2)$ (section 3), conversely $p_X(x) = P_{q+s,q}^{(d)}(\sqrt{1-x})/(2\sqrt{1-x})$. Thus, the identification of the distribution of $X$ from its moments leads ipso facto to that of $R$ and vice versa.

5.2 *The moments* $\langle (1-R^2)^k \rangle_{d,q+s,q}$ *of the walks* $W(d,2,(q+s,q))$

The first representation of $P_{q+s,q}^{(d)}(r)$ (eq. 21) is directly derived from eq. 9 which yields $\langle X^k \rangle_{d,q+s,q} = \langle Y^k \rangle_{q+s,q} \langle Z^k \rangle_d$. The moments $\langle Y^k \rangle_{q+s,q}$ are for instance obtained from eq. B.8 for the mixture of distributions while $\langle Z^k \rangle_d$ is given above eq. 10. Thus:

$$\langle (1-R^2)^k \rangle_{d,q+s,q} = 4^k \times \frac{(q+s)_k (q)_k}{(2q+s)_{2k}} \times \frac{((d-1)/2)_k}{(d-1)_k} \quad (31)$$

It is equivalent to apply eq. 10 to a walk whose step length is distributed according to an asymmetric beta distribution $Be(q+s,q)$ to get eq. 31.

The calculation of the moments from the second representation (eq. 28) makes use of the following relation which mirrors eq. 23 in which powers are replaced by Pochhammer symbols:

$$\frac{(x)_s + (y)_s}{(x+y)_s} = \sum_{j=0}^{\lfloor s/2 \rfloor} (-1)^j C(s,j) \frac{(x)_j (y)_j}{(x+y)_{2j}} \quad (32)$$

$\left((x+y)_s \neq 0, s \geq 2\right)$. As we failed to find eq. 32 in the literature, we derive it in appendix D.

Eqs 27 and 32 confirm readily that $\sum_{j=0}^{\lfloor s/2 \rfloor} (-1)^j w_j = 1$. Eq. 10 applied now to the symmetric walk $W(d, 2, q+j)$ gives:

$$\left\langle (1-R^2)^k \right\rangle_{d,q+j,q+j} = 4^k \frac{(q+j)_k (q+j)_k}{(2q+2j)_{2k}} \times \frac{((d-1)/2)_k}{(d-1)_k} \quad (33).$$

The factor $\dfrac{(q+j)_k (q+j)_k}{(2q+2j)_{2k}}$ can be rewritten as $\dfrac{(q)_k (q)_k}{(2q)_{2k}} \times \dfrac{(2q)_{2j}}{(q)_j (q)_j} \times \dfrac{(q+k)_j (q+k)_j}{(2q+2k)_{2j}}$.

Together with the expression of $w_j$ (eq. 27), the moment $\left\langle (1-R^2)^k \right\rangle_{d,q+s,q}$ becomes:

$$\left\langle (1-R^2)^k \right\rangle_{d,q+s,q} = 4^k \times \frac{(q)_k (q)_k}{(2q)_{2k}} \times \frac{(2q)_s}{2(q)_s} \times \frac{((d-1)/2)_k}{(d-1)_k}$$
$$\times \left\{ \sum_{j=0}^{\lfloor s/2 \rfloor} (-1)^j C(s,j) \frac{(q+k)_j (q+k)_j}{(2q+2k)_{2j}} \right\} \quad (34)$$

The bracketed sum in the right-hand side of eq. 34 is equal to $\dfrac{2(q+k)_s}{(2q+2k)_s}$ (eq. 32). Then:

$$\left\langle (1-R^2)^k \right\rangle_{d,q+s,q} = 4^k \times (q)_k \times \left\{ \frac{(q)_k}{(2q)_{2k}} \times \frac{(2q)_s}{(q)_s} \times \frac{(q+k)_s}{(2q+2k)_s} \right\} \times \frac{((d-1)/2)_k}{(d-1)_k} \quad (35)$$

The bracketed product in eq. 35 simplifies into $\dfrac{(q+s)_k}{(2q+s)_{2k}}$, so that we obtain finally:

$$\left\langle \left(1-R^2\right)^k \right\rangle_{d,q+s,q} = 4^k \times \frac{(q+s)_k (q)_k}{(2q+s)_{2k}} \times \frac{((d-1)/2)_k}{(d-1)_k} \tag{36}$$

which is identical with eq. 31. Without surprise, the previous calculation proves that the densities $P_{q+s,q}^{(d)}(r)$ obtained from eqs 21 and 29 are two representations of the same pdf. It is indeed the permutation invariant length distribution $f_2^{(\Sigma)}(l)$ (eq. 19) associated with the asymmetric beta distribution $Be(q+s,q)$ which is the relevant one as it yields the exact pdf's of the endpoint position and of the final distance.

*5.3 Some examples of random walks $W(d,2,(q+s,q))$*

For $s = 2, 3$, eq. 28 becomes:

$$\begin{cases} P_{q+2,q}^{(d)}(r) = \left(\frac{2q+1}{q+1}\right) P_{q,q}^{(d)}(r) - \left(\frac{q}{q+1}\right) P_{q+1,q+1}^{(d)}(r) \\ P_{q+3,q}^{(d)}(r) = \left(\frac{4q+2}{q+2}\right) P_{q,q}^{(d)}(r) - \left(\frac{3q}{q+2}\right) P_{q+1,q+1}^{(d)}(r) \end{cases} \tag{37}$$

These linear combinations were previously derived by G. Letac (personal communication, 2014). Explicit densities, which are obtained from the pdf's of sections 5.1 and 5.2, are given in appendix E for $s = 2, 3$ where a direct calculation proves that the two representations of $P_{q+s,q}^{(d)}(r)$ are identical for these two values of $s$.

We performed too Monte-Carlo simulations of $W(d,2,(q+s,q))$ random walks for $q = 15/4$ with $s$ ranging between 1 and 6 (figure 1). The beta r.v., $L \sim Be(15/4+s, 15/4)$ (fig. 1a), was simulated with a method described by Devroye (section IX.4 of [48]). As shown above, the asymmetric beta distribution $Be(15/4+s, 15/4)$ and its permutation invariant counterpart $\left(\frac{1}{2} Be(15/4+s, 15/4) + \frac{1}{2} Be(15/4, 15/4+s)\right)$ give rise to identical walks $W(d,2,(15/4+s,15/4))$. In numerical simulations, advantage is then taken from the fact that the simplest distribution to simulate is the former, $Be(15/4+s, 15/4)$, despite the fact that it is

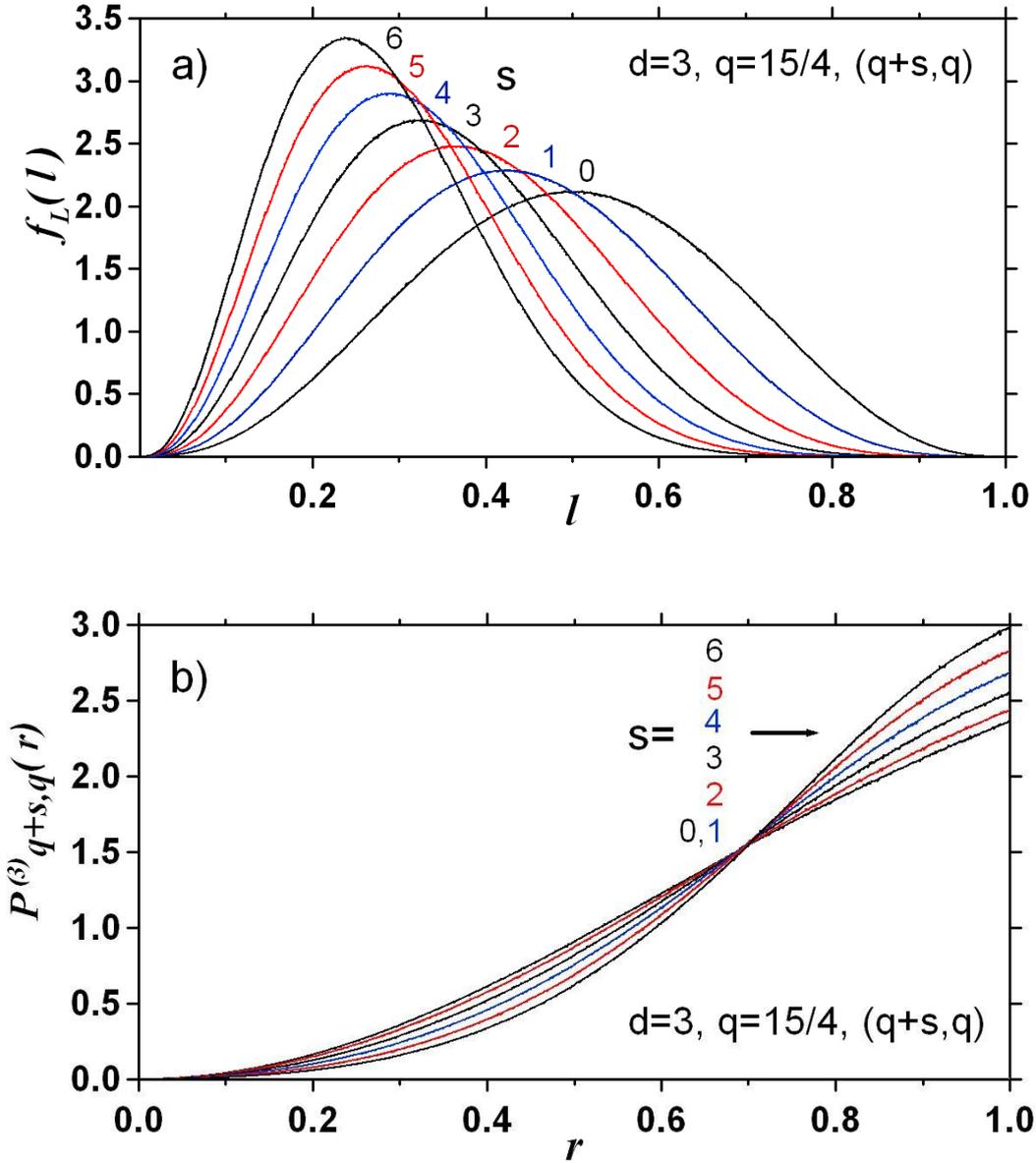

Figure 1: Comparison of the results of Monte-Carlo simulations of $8.10^7$ Dirichlet random walks of two steps in $\mathbb{R}^3$, $W(3,2,(15/4+s,15/4))$, with those calculated for $q=15/4$ and for $s$ varying from 0 to 6 as indicated. The differences between simulated and calculated results are of the order of line thicknesses. Calculated results are obtained from:

a) eq. 2 for the step length pdf $L$, $Be(15/4+s,15/4)$

b) eq. 29 for the pdf of the final distance $R$: $P^{(3)}_{15/4+s,15/4}(r)$ (the pdf's $P^{(3)}_{15/4,15/4}(r)$ $(s=0)$ and $P^{(3)}_{19/4,15/4}(r)$ $(s=1)$ are identical as discussed in section 4.2).

the latter which is the reference distribution as confirmed by fig. 1b which compares the simulated pdf's $P^{(3)}_{15/4+s,15/4}(r)$ $(s=1,..,6)$ to those obtained from the symmetrized distribution (eq. 29). When $s$ becomes larger and larger for a given $q$, the step length distribution $Be(q+s,q)$ concentrates more and more in the vicinity of $l=0$. In consequence, one of the step lengths decreases progressively down to 0 while the other increases concomitantly up to 1. The resulting pdf of the endpoint distance becomes steeper and steeper in the vicinity of 1 as shown by fig. 1.

## 6. Conclusion

First, explicit relations (eqs 12-13) have been given for the pdf's $p^{(d)}_{q,q}(r)$ and $P^{(d)}_{q,q}(r)$ of the endpoint position and of the final distance of a two-step Dirichlet random walk, $W(d,2,q)$, whose associated step length distribution is a symmetric beta distribution $Be(q,q)$. These expressions are valid for any value of $q>0$ and for any $d \geq 2$ such that $q(d)>0$. The specific parameters $q=d/2-1$ and $q=d-1$ of the two families of hyperuniform Dirichlet random walks are seen to emerge quite naturally from the previous pdf's.

Second, $n$-step random walks, whose step length densities are asymmetric, have been considered to conclude that the relevant distributions are actually the permutation invariant step length distributions associated with the initial distributions (eq. 16).

Last, the previous results have been applied to the case of two-step random walks, $W(d,2,(q+s,q))$, with an asymmetric step length distribution $Be(q+s,q)$, where $q$ has any positive value and $s$ is an integer $(s \geq 1)$. Two representations have been derived for the pdf $P^{(d)}_{q+s,q}(r)$ (eqs 21, 29-30), both as sums in a number $1+\lfloor s/2 \rfloor$ of terms. The first one is a sum of non-negative terms and the second is a linear combination, with coefficients of opposite signs, of the pdf's of symmetric Dirichlet walks $W(d,2,q+j)$, $(j=0,..,\lfloor s/2 \rfloor)$.

**Acknowledgments:**

I thank Gérard Letac (University of Toulouse, France) , Emanuele Casini and Andrea Martinelli (University of Como, Italy) for fruitful discussions on asymmetric length distributions and for a critical reading of the first version of the manuscript. Gérard Letac generalized some of the results discussed in section 4 (personal communication).

**Appendix A: Connections between gamma and Dirichlet distributions and between the associated random walks**

A gamma distributed random variable $G$, denoted for brevity as $G \sim \gamma(\alpha, \theta)$, has a probability density function $p_G(x)$ given by [36]:

$$p_G(x) = \frac{x^{\alpha-1} \exp(-x/\theta)}{\theta^\alpha \Gamma(\alpha)} \quad (x > 0) \tag{A.1}$$

where $\alpha > 0$ is the shape parameter and $\theta > 0$ the scale parameter while $\Gamma(\alpha)$ is the Euler gamma function. The characteristic function of $G$ is $\varphi_G(t) = \langle e^{itG} \rangle = 1/(1 - i\theta t)^\alpha$ [36].

A sum $G$ of $n$ independent gamma random variables, $G_i \sim \gamma(q_i, \theta)$ $(i = 1,..,n)$, with identical scale parameters and a priori different shape parameters, is a gamma random variable $G \sim \gamma(n\bar{q}, \theta)$, $\left( n\bar{q} = \sum_{i=1}^{n} q_i \right)$. This is readily deduced from the characteristic function of the sum $G = \sum_{k=1}^{n} G_k$, $\varphi_G(t) = \langle e^{itG} \rangle = 1 / \left( \prod_{k=1}^{n} (1 - it\theta)^{q_k} \right) = 1/(1 - it\theta)^{n\bar{q}}$. As the scale parameter is irrelevant in the present context, its value will be fixed at 1 from now on.

The Dirichlet distribution can be obtained [36] from a set of $n = m+1$ independent gamma random variables $G_i \sim \gamma(q_i, 1)$ $(i = 1,...,n)$ by defining $G = \sum_{j=1}^{n} G_j \sim \gamma(n\bar{q}, 1)$ and $L_j = G_j/G$ $(j = 1,..,n)$. The distribution of $\mathbf{L}_{(n)} = (L_1, L_2,..., L_n)$ is then a Dirichlet distribution with parameters $\mathbf{q}_{(n)} = (q_1,..., q_n)$, $\mathbf{L}_{(n)} \sim D(\mathbf{q}_{(n)})$, and a pdf ([36], p. 17):

$$\begin{cases} f(l_1,..,l_m) = K_n \prod_{i=1}^{n} l_i^{q_i - 1} \\ l_n = 1 - \sum_{i=1}^{m} l_i, \quad l_i > 0, \ i = 1,..,n \end{cases} \tag{A.2}$$

where $K_n = \Gamma(n\bar{q}) \big/ \left( \prod_{i=1}^{n} \Gamma(q_i) \right)$. If the $n$ components of $\boldsymbol{L}_{(n)} \sim D(\boldsymbol{q}_{(n)})$ are collected into $k$ groups and summed up to form $k$ new components $\boldsymbol{S}_{(k)} = (S_1,..,S_k), \left( \sum_{i=1}^{k} S_i = 1 \right)$, then the distribution of $\boldsymbol{S}_{(k)}$ is $D\left(\boldsymbol{q}^*_{(k)}\right)$ where each $q^*_i$ $(i=1,..,k)$ is the sum of the parameters $q_j$'s of the components of $\boldsymbol{L}_{(n)}$ which add up to $S_i$. This amalgamation property [36] results directly from the characteristics of the sum of independent gamma random variables described above. The marginal distribution of any component $L_k$ $(k=1,..,n)$ is then obtained from the two components $S_1 = L_k$ and $S_2 = \sum_{i=1, i \neq k}^{n} L_i = 1 - S_1$. The marginal distribution $f_k(l)$ of $L_k$ is thus a beta distribution whose pdf is [36]:

$$f_k(l) = \frac{l^{q_k-1}(1-l)^{n\bar{q}-q_k-1}}{B(q_k, n\bar{q}-q_k)} \quad l \in [0,1] \quad (A.3)$$

We consider the stochastic relation between the $n$-dimensional random vectors $\boldsymbol{G}_{(n)}$ and $\boldsymbol{L}_{(n)}$ (section 4 of [26]):

$$\boldsymbol{G}_{(n)} \triangleq S\boldsymbol{L}_{(n)} \quad (A.4)$$

where $A \triangleq B$ means that the random variables $A$ and $B$ are identically distributed, $S$ is gamma distributed, $\gamma(nq,1)$, $\boldsymbol{L}_{(n)} = (L_1, L_2,..,L_n)$ is Dirichlet distributed, $D(\boldsymbol{q}_{(n)} = (q,q,..q))$ and $S$ and $\boldsymbol{L}_{(n)}$ are independent. Then the vector $\boldsymbol{G}_{(n)} = (G_1 \triangleq SL_1, G_2 \triangleq SL_2,..,G_n \triangleq SL_n)$ has independent $\gamma(q,1)$ components ([36], p. 148). From $\boldsymbol{G}_{(n)}$, we define a gamma random walk $G(d,n,q)$ whose endpoint position is:

$$\boldsymbol{R}_G = \sum_{i=1}^{n} G_i \boldsymbol{U}_i^{(d)} \triangleq S\left( \sum_{i=1}^{n} L_i \boldsymbol{U}_i^{(d)} \right) \quad (A.5)$$

where $U_i^{(d)}$ $(i=1,..,n)$ are $n$ independent and identically distributed unit vectors. Similarly $L_{(n)}$ defines a Dirichlet random walk $W(d,n,q)$ whose endpoint position is:

$$R = \sum_{i=1}^{n} L_i\, U_i^{(d)} \qquad (A.6)$$

The connection between the endpoint positions $R_G$ and $R$ of the random walks $G(d,n,q)$ and $W(d,n,q)$, which results from eq. A.4, can then be condensed in the following stochastic representation:

$$R_G \triangleq SR \qquad (A.7)$$

where $S$ and $R$ are independent. We denote the pdf's of the endpoint position and of the endpoint distance of the gamma walk $G(d,n,q)$ respectively as $g_{q_{(n)}}^{(d)}(r)$ and $G_{q_{(n)}}^{(d)}(r)$. Translating eq. A.7 in terms of pdf's, $g_{q_{(n)}}^{(d)}(r)$ is related to the pdf of the endpoint position $p_{q_{(n)}}^{(d)}(r)$ of the Dirichlet walk $W(d,n,q)$ by (section 4 of [26]):

$$g_{q_{(n)}}^{(d)}(r) = \frac{r^{nq-d}}{\Gamma(nq)} \int_{1}^{\infty} \exp(-rt)\, t^{nq-d-1}\, p_{q_{(n)}}^{(d)}\!\left(\frac{1}{t}\right) dt \qquad (A.8)$$

with $p_{q_{(n)}}^{(d)}(r)=0$ for $r>1$. The pdf $g_{q_{(n)}}^{(d)}(r)$ is thus the Laplace transform of $t^{nq-d-1} p_{q_{(n)}}^{(d)}(1/t)$. The known pdf $g_{q_{(n)}}^{(d)}(r)$ of the walk $G(d,n,d)$ was used in [26] to derive $p_{q_{(n)}}^{(d)}(r)$ for the walk $W(d,n,d)$. Finally, $G_{q_{(n)}}^{(d)}(r)$ is given by $G_{q_{(n)}}^{(d)}(r) = \frac{2\pi^{d/2}}{\Gamma(d/2)} r^{d-1} g_{q_{(n)}}^{(d)}(r)$ (eq. 4).

**Appendix B: Distribution of** $Y = 4L(1-L)$ **for** $L \sim Be(q_1, q_2)$

The continuous r.v.'s $L$ and $Y$ have a common support, $[0,1]$. When $n = 2$, the Dirichlet distribution (eq. 1) reduces to a beta distribution, $L \sim Be(q_1, q_2)$ (eq. 2). From the relation between probabilities:

$$P(Y \leq y) = P\left(L \leq \frac{1}{2} - \frac{\sqrt{1-y}}{2}\right) + P\left(L > \frac{1}{2} + \frac{\sqrt{1-y}}{2}\right) \quad y \in [0,1] \tag{B.1}$$

and from $\left|\dfrac{dl}{dy}\right| = \dfrac{1}{4\sqrt{1-y}}$, we get the pdf of $Y$:

$$p_Y(y) = \left\{ f_L\left(\frac{1}{2} - \frac{\sqrt{1-y}}{2}\right) + f_L\left(\frac{1}{2} + \frac{\sqrt{1-y}}{2}\right) \right\} \times \frac{1}{4\sqrt{1-y}} \tag{B.2}$$

When $q_1 = q_2 = q$, $f_L(l) = l^{q-1}(1-l)^{q-1}/B(q,q)$, eq. B.2 becomes, after applying the duplication formula of the gamma function, $\Gamma(2q) = \dfrac{4^q \Gamma(q)\Gamma(q+1/2)}{2\sqrt{\pi}}$:

$$p_Y(y) = \frac{2}{B(q,q)} \times \left[\left(\frac{1}{2} - \frac{\sqrt{1-y}}{2}\right)\left(\frac{1}{2} + \frac{\sqrt{1-y}}{2}\right)\right]^{q-1} \times \frac{1}{4\sqrt{1-y}} = \frac{y^{q-1}(1-y)^{-1/2}}{B(q,1/2)} \tag{B.3}$$

which shows that $Y$ has a beta distribution, $Y \sim Be\left(q, \dfrac{1}{2}\right)$ [29].

When the distribution of $L$ is asymmetric, $q_1 \neq q_2$, $f_L(l) = l^{q_1-1}(1-l)^{q_2-1}/B(q_1,q_2)$, we define $q = \min(q_1,q_2)$, $Q = \max(q_1,q_2)$ and $\Delta q = Q - q = |q_1 - q_2|$, to write:

$$p_Y(y) = \frac{y^{q-1}(1-y)^{-1/2}}{4^q B(q_1,q_2)} \times \left\{\left(\frac{1}{2} - \frac{\sqrt{1-y}}{2}\right)^{\Delta q} + \left(\frac{1}{2} + \frac{\sqrt{1-y}}{2}\right)^{\Delta q}\right\} \tag{B.4}$$

When $\Delta q > 1$, the pdf's of $Y$ (eq. B.4) and consequently the pdf's of the final distance $R$ (eqs 8 and 9), are seen to be identical when obtained either from the length distribution, $L \sim Be(q_1, q_2)$, or from the symmetric mixture, $L \sim \frac{1}{2} Be(q_1, q_2) + \frac{1}{2} Be(q_2, q_1)$. This point is further discussed in section 4.

When $\Delta q = s$, where $s$ is a positive integer, the distribution of $Y$ becomes a mixture of beta distributions. It reduces to a beta distribution, $Y \sim Be\left(q, \frac{1}{2}\right)$, for $s=1$ (see too section 4.2). Using the fact that:

$$\left(\frac{1}{2} - \frac{\sqrt{1-y}}{2}\right)^s + \left(\frac{1}{2} + \frac{\sqrt{1-y}}{2}\right)^s = 2^{1-s}\left\{\sum_{k=0}^{\lfloor s/2 \rfloor} \binom{s}{2k} (1-y)^k\right\} \quad (B.5)$$

the distribution of $Y$ (eq. B.4) becomes indeed:

$$\begin{cases} p_Y(y) = \sum_{k=0}^{\lfloor s/2 \rfloor} c_k \times \left\{\dfrac{y^{q-1}(1-y)^{k-1/2}}{B(q, k+1/2)}\right\} \\ c_k = \binom{s}{2k} \times \dfrac{B(q, k+1/2)}{2^{2q+s-1} B(q+s, q)} \end{cases} \quad (B.6)$$

It is a mixture of $1 + \left\lfloor \dfrac{s}{2} \right\rfloor$ beta distributions, $p_Y(y) = \sum_{k=0}^{\lfloor s/2 \rfloor} c_k p_{Y_k}(y)$ with $Y_k \sim Be(q, k+1/2)$. As only normalized distributions have been dealt with all along the calculation, it follows that $\sum_{k=0}^{\lfloor s/2 \rfloor} c_k = 1$. Equivalently, the latter relation writes:

$$\sum_{k=0}^{\lfloor s/2 \rfloor} \binom{s}{2k} \times B(q, k+1/2) = 2^{2q+s-1} B(q+s, q) \quad (B.7)$$

A direct proof of eq. B.7 is straightforward, being the converse of the previous calculation. The definition of the beta function (eq. 2), the use of eq. B.5 and a change of variable $x = \sqrt{1-t}$ gives indeed:

$$\sum_{k=0}^{\lfloor s/2 \rfloor} \binom{s}{2k} \times B(q, k+1/2) = \int_0^1 t^{q-1} (1-t)^{-1/2} \left\{ \sum_{k=0}^{\lfloor s/2 \rfloor} \binom{s}{2k} \times (1-t)^k \right\} dt =$$

$$= \int_{-1}^1 (1+x)^{q-1} (1-x)^{q+s-1} dx = 2^{2q+s-1} B(q+s, q)$$

where the last equality results from integral 3.196.3 of [42]. Similarly the moments $\langle Y^k \rangle_{q+s,q}$ are easily retrieved from the mixture of distributions, using eqs 2, B.6 and B.7:

$$\langle Y^k \rangle_{q+s,q} = \sum_{n=0}^{\lfloor s/2 \rfloor} c_n \langle Y_n^k \rangle_{q+s,q} = \sum_{n=0}^{\lfloor s/2 \rfloor} \binom{s}{2n} \times \frac{B(q, n+1/2)}{2^{2q+s-1} B(q+s, q)} \times \frac{(q)_k}{(q+n+1/2)_k} =$$

$$= \sum_{n=0}^{\lfloor s/2 \rfloor} \binom{s}{2n} \times \frac{B(q+k, n+1/2)}{2^{2q+s-1} B(q+s, q)} = \frac{4^k B(q+s+k, q+k)}{B(q+s, q)} = \frac{4^k (q+s)_k (q)_k}{(2q+s)_{2k}} \quad \text{(B.8)}$$

which is the contribution of $\langle Y^k \rangle_{q+s,q}$ in eq. 10 when $L \sim Be(q+s, q)$.

**Appendix C: Distribution of the product of two independent beta r.v.'s**

Let us consider a r.v. $X$ which is the product of two independent r.v.'s $Y$ and $Z$ which are distributed according to beta distributions, respectively $Y \sim Be(\alpha_1, \beta_1)$ and $Z \sim Be(\alpha_2, \beta_2)$. Then the pdf of $X$ is given by [38-41]:

$$p_X(x) = \frac{B(\beta_1, \beta_2)}{B(\alpha_1, \beta_1) B(\alpha_2, \beta_2)} \times x^{\alpha_1 - 1} (1-x)^{\beta_1 + \beta_2 - 1} {}_2F_1(\alpha_1 + \beta_1 - \alpha_2, \beta_2; \beta_1 + \beta_2; 1-x) \quad \text{(C.1)}$$

($x \in [0,1]$), where ${}_2F_1(a, b; c; x)$ is a Gauss hypergeometric function. The moments $\langle X^k \rangle$, where $k$ is chosen here to be a positive integer, are simply obtained from the product of the corresponding moments of $Y$ and $Z$ given both by eq. 2:

$$\langle X^k \rangle = \langle Y^k \rangle \langle Z^k \rangle = \frac{(\alpha_1)_k (\alpha_2)_k}{(\alpha_1 + \beta_1)_k (\alpha_2 + \beta_2)_k} \quad \text{(C.2)}$$

**Appendix D: The relation** $\sum_{j=0}^{\lfloor n/2 \rfloor}(-1)^j C(n,j)\dfrac{(x)_j (y)_j}{(x+y)_{2j}} = \dfrac{(x)_n + (y)_n}{(x+y)_n}$ (eq. 32)

Relations, in which powers are "replaced" by Pochhammer symbols, have been known for a long time. This is for instance the case for the expansion of the power of a binomial whose analog in term of Pochhammer symbols is the Vandermonde's identity $(x+y)_n = \sum_{k=0}^{n}\binom{n}{k}(x)_k (y)_{n-k}$ (see among others [37]). The relation we consider is similarly analogous to a polynomial identity, $x^n + y^n = \sum_{j=0}^{\lfloor n/2 \rfloor}(-1)^j C(n,j)(xy)^j (x+y)^{n-2j}$ (eqs 22 and 23). Robbins [44] mentions that the latter identity would be due to Lucas and possibly to Lagrange while Gould [43] shows that it is a special case of a formula first established by Girard in 1629 and later given by Waring in the eighteen century. The validity of eq. 32 is proven here by a method which is likely only one among many others and is in no way claimed to be the simplest one. Another example of a pair of analogous relations is given by eqs D.7 and D.9 below.

Here, we consider the case where $n$ is an integer larger than 1 as the relation reduces to 1=1 for $n=1$. We assume further that $(x+y)_n \neq 0$. We define first the sum:

$$\begin{cases} S_n(x,y) = 1 + \sum_{j=1}^{\lfloor n/2 \rfloor}(-1)^j C(n,j)\dfrac{(x)_j (y)_j}{(x+y)_{2j}} \\ C(n,j) = \dfrac{n}{(n-j)}\binom{n-j}{j} = \dfrac{n\Gamma(n-j)}{\Gamma(n+1-2j)\, j!} \end{cases} \quad (D.1)$$

where the $C(n,j)$ are the Lucas coefficients [45]. Because $j \leq \left\lfloor \dfrac{n}{2} \right\rfloor$, the Pochhammer symbols $(1-n)_j$ and $(-n)_{2j}$ are different from zero for $n>1$. The falling factorial is defined as:

$$(a)^{(j)} = \Gamma(a+1)/\Gamma(a+1-j) = a(a-1)\ldots(a+1-j) = (-1)^j (-a)_j \quad (D.2)$$

Thus:

$$\begin{cases} \Gamma(n-j) = \Gamma(n)/(n-1)^{(j)} = (-1)^j \Gamma(n)/(1-n)_j \\ \Gamma(n+1-2j) = \Gamma(n+1)/(-n)_{2j} = \Gamma(n+1)/\left(4^j(-n/2)_j((1-n)/2)_j\right) \end{cases} \quad (D.3)$$

where the duplication formula (eq. 2) has been applied to $(-n)_{2j}$. Finally, applying again the duplication formula to $(x+y)_{2j}$, eq. D.1 becomes:

$$S_n(x,y) = 1 + \sum_{j=1}^{\lfloor n/2 \rfloor} \frac{(-n/2)_j((1-n)/2)_j}{(1-n)_j} \times \frac{(x)_j(y)_j}{\left(\frac{x+y}{2}\right)_j \left(\frac{x+y+1}{2}\right)_j} \times \frac{1}{j!} \quad (D.4)$$

The latter relation can be expressed in term of a generalized hypergeometric function:

$$S_n(x,y) = {}_4F_3\left(-\frac{n}{2}, \frac{1-n}{2}, x, y; 1-n, \frac{x+y}{2}, \frac{x+y+1}{2}; 1\right) \quad (D.5)$$

Prudnikov et al. don't give the value of the latter hypergeometric function but they give instead the following value (7.5.3.57 of [46]):

$$_4F_3\left(-\frac{n}{2}, \frac{1-n}{2}, x, y; -n, \frac{x+y}{2}+1, \frac{x+y+1}{2}; 1\right) = \frac{(x)_{n+1} - (y)_{n+1}}{(x-y)(x+y+1)_n} \quad (D.6)$$

Reversing the method used above, eq. D.6 can be converted back into the following sum, which involves Pochhammer symbols, after dividing each member by $(x+y)$:

$$\sum_{j=0}^{\lfloor n/2 \rfloor} (-1)^j \binom{n-j}{j} \frac{(x)_j(y)_j}{(x+y)_{2j+1}} = \frac{(x)_{n+1} - (y)_{n+1}}{(x-y)(x+y)_{n+1}} \quad (D.7)$$

The polynomial identity given as eq. 22 of [43] is:

$$\sum_{j=0}^{\lfloor n/2 \rfloor}(-1)^j \binom{n-j}{j}(x+y)^{n-2j}(xy)^j = \frac{x^{n+1}-y^{n+1}}{(x-y)} \qquad (D.8)$$

Interestingly, eq. D.7 is then seen to be the analogous of the following identity obtained by dividing each member of eq. D.8 by $(x+y)^{n+1}$:

$$\sum_{j=0}^{\lfloor n/2 \rfloor}(-1)^j \binom{n-j}{j}\frac{x^j y^j}{(x+y)^{2j+1}} = \frac{x^{n+1}-y^{n+1}}{(x-y)(x+y)^{n+1}} \qquad (D-9)$$

Back to eq. D.6, we use first a relation between contiguous hypergeometric functions [47]:

$$\begin{aligned}
&\alpha(\gamma-\delta)\,_4F_3(\alpha+1,\alpha_2,\alpha_3,\alpha_4;-n,\gamma+1,\delta+1,\beta_3;z) \\
&-\delta(\gamma-\alpha)\,_4F_3(\alpha,\alpha_2,\alpha_3,\alpha_4;-n,\gamma+1,\delta,\beta_3;z) \\
&+\gamma(\delta-\alpha)\,_4F_3(\alpha,\alpha_2,\alpha_3,\alpha_4;-n,\gamma,\delta+1,\beta_3;z)=0
\end{aligned} \qquad (D.10)$$

With $\alpha = -\frac{n}{2}, \alpha_2 = \frac{1-n}{2}, \alpha_3 = x, \alpha_4 = y, \gamma = -n, \delta = \frac{x+y}{2}, \beta_3 = \frac{x+y+1}{2}$, eq. D.10 becomes:

$$\begin{aligned}
&(x+y+2n)\,_4F_3\left(\frac{1-n}{2}, 1-\frac{n}{2}, x, y; 1-n, \frac{x+y}{2}+1, \frac{x+y+1}{2}; 1\right) \\
&+(x+y)\,_4F_3\left(\frac{1-n}{2}, -\frac{n}{2}, x, y; 1-n, \frac{x+y}{2}, \frac{x+y+1}{2}; 1\right) \\
&-2(x+y+n)\,_4F_3\left(\frac{1-n}{2}, -\frac{n}{2}, x, y; -n, \frac{x+y}{2}+1, \frac{x+y+1}{2}; 1\right)=0
\end{aligned} \qquad (D.11)$$

From relation D.6, the first and the third hypergeometric functions in eq. D.11 are respectively equal to $\frac{(x)_n-(y)_n}{(x-y)(x+y+1)_{n-1}}$ and $\frac{(x)_{n+1}-(y)_{n+1}}{(x-y)(x+y+1)_n}$ while the second is the hypergeometric function we wish to express. Therefore:

$$_4F_3\left(\frac{1-n}{2}, -\frac{n}{2}, x, y; 1-n, \frac{x+y}{2}, \frac{x+y+1}{2}; 1\right) =$$

$$= \frac{2(x+y+n)\left((x)_{n+1} - (y)_{n+1}\right)}{(x-y)(x+y)(x+y+1)_n} - \frac{(x+y+2n)\left((x)_n - (y)_n\right)}{(x-y)(x+y)(x+y+1)_{n-1}} =$$

$$= \frac{1}{(x-y)(x+y)_n}\left[2(x)_{n+1} - 2(y)_{n+1} - (x+y+2n)\left((x)_n - (y)_n\right)\right] \qquad (D.12)$$

Writing now:

$$-(x+y+2n)\left((x)_n - (y)_n\right) = -(x+n+y+n)\left((x)_n - (y)_n\right) =$$
$$= -(x)_{n+1} + (y)_{n+1} + (x-y+y+n)(y)_n - (y-x+x+n)(x)_n =$$
$$= -2(x)_{n+1} + 2(y)_{n+1} + (x-y)\left((x)_n + (y)_n\right)$$

we obtain finally the relation sought for from eqs D.5 and D.12:

$$\sum_{j=0}^{\lfloor n/2 \rfloor} (-1)^j C(n,j) \frac{(x)_j (y)_j}{(x+y)_{2j}} = \frac{(x)_n + (y)_n}{(x+y)_n} \qquad (D.13)$$

When $x = y$, the hypergeometric function (eq. D.5) becomes $_3F_2\left(-\frac{n}{2}, \frac{1-n}{2}, x; 1-n, x+\frac{1}{2}; 1\right)$ which can then be obtained from relation 7.4.4.113 of [46], $_3F_2\left(-\frac{n}{2}, \frac{1-n}{2}, x; b, \frac{3}{2}+x-b-n; 1\right) = \frac{(2b-2x-1)_n (2b+n-1)_n}{4^n (b-x-1/2)_n (b)_n}$. This leads in turn to eq. D.13 with $x = y$.

**Appendix E: The pdf's** $P_{q+2,q}^{(d)}(r)$ **and** $P_{q+3,q}^{(d)}(r)$ **(eq. 37)**

For the first representation and for $s = 2$, eq. 21 gives:

$$P_{q+2,q}^{(d)}(r) = \frac{2^{d-2-2q}}{B(q+2,q)} \times r^{d-1}(1-r^2)^{q-1} \times$$

$$\left\{ {}_2F_1\left(q+1-\frac{d}{2}, \frac{d-1}{2}; \frac{d}{2}; r^2\right) + \left(\frac{r^2}{d}\right) \times {}_2F_1\left(q+2-\frac{d}{2}, \frac{d-1}{2}; \frac{d}{2}+1; r^2\right) \right\} \quad (E.1)$$

while for $s = 3$:

$$P_{q+3,q}^{(d)}(r) = \frac{2^{d-3-2q}}{B(q+3,q)} \times r^{d-1}(1-r^2)^{q-1} \times$$

$$\left\{ {}_2F_1\left(q+1-\frac{d}{2}, \frac{d-1}{2}; \frac{d}{2}; r^2\right) + \left(\frac{3r^2}{d}\right) \times {}_2F_1\left(q+2-\frac{d}{2}, \frac{d-1}{2}; \frac{d}{2}+1; r^2\right) \right\} \quad (E.2)$$

For the second representation, eqs 12 and 37 yield respectively:

$$P_{q+2,q}^{(d)}(r) = 2^{d-1} r^{d-1}(1-r^2)^{q-1} \times$$

$$\times \left\{ \begin{array}{l} \frac{(2q+1)}{(q+1)B(q,1/2)} \times {}_2F_1\left(q+1-\frac{d}{2}, \frac{d-1}{2}; \frac{d}{2}; r^2\right) \\ -\frac{q}{(q+1)B(q+1,1/2)} \times (1-r^2) \times {}_2F_1\left(q+2-\frac{d}{2}, \frac{d-1}{2}; \frac{d}{2}; r^2\right) \end{array} \right\} \quad (E.3)$$

and

$$P_{q+3,q}^{(d)}(r) = 2^{d-1} r^{d-1}(1-r^2)^{q-1} \times$$

$$\times \left\{ \begin{array}{l} \frac{(4q+2)}{(q+2)B(q,1/2)} \times {}_2F_1\left(q+1-\frac{d}{2}, \frac{d-1}{2}; \frac{d}{2}; r^2\right) \\ -\frac{3q}{(q+2)B(q+1,1/2)} \times (1-r^2) \times {}_2F_1\left(q+2-\frac{d}{2}, \frac{d-1}{2}; \frac{d}{2}; r^2\right) \end{array} \right\} \quad (E.4)$$

The hypergeometric function $_2F_1\left(q+2-\dfrac{d}{2},\dfrac{d-1}{2};\dfrac{d}{2}+1;r^2\right)$ is then expressed from a Gauss'relation for contiguous hypergeometric functions (eq. 15.2.20 of [49]) as:

$$_2F_1\left(q+2-\dfrac{d}{2},\dfrac{d-1}{2};\dfrac{d}{2}+1;r^2\right)=$$
$$\dfrac{d}{r^2}\times\left\{_2F_1\left(q+1-\dfrac{d}{2},\dfrac{d-1}{2};\dfrac{d}{2};r^2\right)-\left(1-r^2\right)\times\,_2F_1\left(q+2-\dfrac{d}{2},\dfrac{d-1}{2};\dfrac{d}{2};r^2\right)\right\} \quad (E.5)$$

From $\dfrac{1}{B(q',q')}=\dfrac{2^{2q'-1}}{B(q',1/2)}$ we obtain:

$$\dfrac{1}{B(q+2,q)}=\left(\dfrac{2q+1}{q+1}\right)\times\dfrac{2^{2q}}{B(q,1/2)}=\left(\dfrac{q}{q+1}\right)\times\dfrac{2^{2q+1}}{B(q+1,1/2)} \quad (E.6)$$

It suffices now to insert eq. E.5 into eq. E.1 and to use eq. E.6 to transform the first representation of the pdf $P^{(d)}_{q+2,q}(r)$ (eq. E.1) into the second (eq. E.3). Similarly:

$$\dfrac{1}{B(q+3,q)}=\left(\dfrac{4q+2}{q+2}\right)\times\dfrac{2^{2q}}{B(q,1/2)}=\left(\dfrac{q}{q+2}\right)\times\dfrac{2^{2q+2}}{B(q+1,1/2)} \quad (E.7)$$

The left hypergeometric function on the right-hand side of eq. E.2 is multiplied by four once eq. E.5 has been inserted into eq. E.2. Then, eq. E.7 transforms the first representation of the pdf $P^{(d)}_{q+3,q}(r)$ (eq. E.2) into the second (eq. E.4). The two representations of $P^{(d)}_{q+s,q}(r)$ (sections 5.1 and 5.2) are thus proven by a direct calculation to be identical for $s=2,3$.